
\input harvmac
\def\321{$SU(3) \times SU(2) \times U(1)$}
\def\btheta{\bar \theta}
\def\dtwoth{d^4 x d^2 \theta\ }
\def\dfourth{d^4 x d^2 \theta d^2 \btheta\ }
\def\btheta{\bar \theta}
\def\E{{\cal E}}
\def\L{{\cal L}}

\catcode`\@=11 
\def\lsim{\mathrel{\mathpalette\@versim<}}
\def\gsim{\mathrel{\mathpalette\@versim>}}
\def\@versim#1#2{\vcenter{\offinterlineskip
      \ialign{$\m@th#1\hfil##\hfil$\crcr#2\crcr\sim\crcr } }}
\catcode`\@=12 

\Title{\vbox{
\hbox{JHU-TIPAC-93018}
\hbox{hep-ph/9307317}}}
{\vbox{\centerline{Destabilizing Divergences in Supergravity-Coupled}
\vskip2pt \centerline{Supersymmetric Theories$^*$}}}
\footnote{}{${}^*$\ \ This work was supported in part by the U.S.
National Science Foundation, grant PHY90-96198, and by the Texas
National Research Laboratory Commission, grant RGFY93-292.}
\vskip.5cm

\centerline{Jonathan Bagger and Erich Poppitz}
\centerline{\it Department of Physics and Astronomy}
\centerline{\it The Johns Hopkins University}
\centerline{\it Baltimore, MD\  21218}

\vskip 3.0truecm
{\centerline {\bf Abstract}}
\noindent
Nonrenormalizable couplings in supergravity-coupled
supersymmetric theories can give rise to power-law
divergences that destabilize the weak-scale hierarchy.
For the case of the standard-model gauge group, the
problem can arise in theories with \321 gauge-singlet
chiral superfields.  The minimal supersymmetric
standard model is free from such destabilizing
divergences.

\Date{July, 1993}

\newsec{Introduction}

Low-energy supersymmetry is motivated by the fact that
in the standard model, the Higgs mass $M_H$ receives
quadratically-divergent radiative corrections.  If these
divergences are cut off at a scale $\Lambda$, representing
the scale of new physics, the corrections are of order
$\delta M^2_H \sim \Lambda^2$.  For $\Lambda^2 \gg M^2_H$, a
dramatic fine tuning is needed to keep $M_H$ near the
weak scale $M_W$.  This is known as the naturalness
problem of the standard model \ref{\hierarchy}{S.~Weinberg,
Phys. Rev. D13 (1976) 974; D19 (1979) 1277\semi
L.~Susskind, Phys. Rev. D20 (1979) 2619\semi
G.~{}'t Hooft, {\it in} Recent Developments in Gauge Theories,
ed. G.~{}'t Hooft, {\it et al.} (Plenum, 1980).}.

Supersymmetric theories cure the naturalness problem by
introducing new particles in just such a way as to cancel
the quadratic divergences.  Assuming that the new particles
have masses of order $M_W$, the hierarchy $M^2_H \ll
\Lambda^2$ is preserved by radiative corrections.
Supersymmetry renders the hierarchy technically natural
\hierarchy.

Fortunately, complete supersymmetry invariance is not
necessary for the cancellation of quadratic divergences.
For example, it is possible for supersymmetry to be
spontaneously broken, with an order parameter $\langle
F \rangle \sim M^2_W$, and still preserve the
hierarchy \ref{\Witten}{E.~Witten, Nucl. Phys. B185
(1981) 513.}.

It is also possible to break supersymmetry explicitly,
but softly, and not reintroduce quadratic divergences
\ref{\DimGeorgiSakai}{S.~Dimopoulos and H.~Georgi, Nucl.
Phys. B193 (1981) 150\semi
N.~Sakai, Zeit. Phys. C11 (1981) 153.}.
Indeed, this is the approach that is often taken in
constructing realistic models.  For this procedure to
work, the soft mass parameters must be of order the weak
scale $M_W$.

Supersymmetry, however, is a spacetime symmetry, so it
does not make sense to break it explicitly.  Nevertheless,
an effective theory with explicit soft breakings can
be found by assuming the existence of hidden matter
that couples only gravitationally to the visible world.
If supersymmetry is spontaneously broken in the hidden
sector, with an order parameter $\langle F \rangle \sim
M_W M_P$ (where $M_P$ is the Planck mass), soft breakings
with mass parameters of order $M_W$
are induced in the visible world \ref{\grav}
{H.~Nilles, Phys. Lett. 115B (1982) 193\semi
A.~Chamseddine, R.~Arnowitt and P.~Nath, Phys. Rev.
Lett. 49 (1982) 970\semi
R.~Barbieri, S.~Ferrara and C.~Savoy, Phys. Lett.
119B (1982) 343\semi
L.~Iba\~nez, Nucl. Phys. B218 (1982) 514\semi
H.~Nilles, M.~Srednicki and D.~Wyler, Phys. Lett.
120B (1983) 346\semi
E.~Cremmer, P.~Fayet and L.~Girardello, Phys. Lett.
122B (1983) 41\semi
L.~Hall, J.~Lykken and S.~Weinberg, Phys. Rev. D27
(1983) 2359\semi
L.~Alvarez-Gaum\'e, J.~Polchinski and M.~Wise, Nucl.
Phys. B221 (1983) 495.}.

In the limit $M_P \rightarrow \infty$, this process
gives an effective theory that
is renormalizable and supersymmetric, up to certain
soft breakings that preserve the hierarchy $M^2_H \ll
\Lambda^2$.  However, we know that in the real world,
$M_P \ne \infty$, and that gravitational effects give
rise to an effective Lagrangian with a tower of
nonrenormalizable operators, suppressed by powers
of $M_P$.  In this letter we will show that these
nonrenormalizable operators can destabilize the
hierarchy in supersymmetric models with singlet
superfields \ref{\Ellwanger}{U.~Ellwanger, Phys. Lett.
133B (1983) 187.}.

Most previous studies of singlet superfields have been
done in the context of grand unification, where the
couplings of singlets to superheavy particles can
destabilize the weak-scale hierarchy \ref{\gutsoft}
{J.~Polchinski and
L.~Susskind, Phys. Rev. D26 (1982) 3661\semi
H.~Nilles, M.~Srednicki and D.~Wyler, Phys. Lett. 124B
(1983) 337\semi
A.~Lahanas, Phys. Lett. 124B (1983) 341.}.
We shall see that destabilizing divergences are a
general feature of supergravity models with singlet
superfields, even those without any superheavy
particles or unification scheme \Ellwanger.  We shall
show that naturalness requires the mass of the lightest
singlet to be greater than $\Lambda^2/M_P$,
where $\Lambda \lsim M_P$ is the scale where
the effective theory breaks down.

\newsec{Renormalizable Couplings}

We will begin our analysis by recalling the pertinent
features of supersymmetric theories in superspace.
We use the superspace notation because it simplifies
power counting in supersymmetric theories.

In supersymmetric theories, all matter fields are
described in terms of chiral and
vector superfields.  For the purposes of this letter,
we will focus our attention on the supersymmetric standard
model (and its extensions) with gauge group
$G = SU(3) \times SU(2) \times U(1)$.
We will ignore the vector superfields, and restrict our
attention to the chiral Higgs superfields $H_1$ and $H_2$, as
well as a singlet chiral superfield $N$.  (We assume $N$
to be a singlet under the \321 gauge group, as well as
under any discrete symmetries that are preserved by gravity.)
In the chiral basis, these superfields have the following
component-field expansions \ref{\WessBagger}{We follow the
conventions of J.~Wess and J.~Bagger, \it Supersymmetry and
Supergravity, \rm (Princeton, 1992).}:
\eqn{\component}{\eqalign{
H_1\ &=\ h_1 + \theta \chi_1 + \theta\theta F_1 \cr
H_2\ &=\ h_2 + \theta \chi_2 + \theta\theta F_2 \cr
N\ &=\ n + \theta \chi_n + \theta\theta F_n\ . \cr}}

The Lagrangian for these fields contains two types of
terms:  $D$-terms
\eqn{\dfour}{\L_D\ =\ \int \dfourth ( N^+ N + H_1^+ H_1 + H_2^+ H_2 )\ ,}
and $F$-terms
\eqn{\dtwo}{\L_F\ =\ \int \dtwoth ( M_N N^2 + \lambda N^3 +
\mu H_1 H_2 + g N H_1 H_2 )\ +\ h.c.\ ,}
where any linear term in $N$ is removed by shifting, $\mu \sim
M_W$, and $M_N$ is less than the cutoff $\Lambda$.
If supersymmetry is unbroken, the $F$-terms are not
renormalized, while the $D$-terms are renormalized at
most logarithmically.  The Lagrangians \dfour\ and
\dtwo\ are invariant under supersymmetry and rigid \321.

Let us now suppose that the Lagrangian $\L_D + \L_F$
describes part of the visible world in a
supergravity-coupled supersymmetric
theory with a singlet superfield $N$.  The
supergravity coupling of this Lagrangian is simple to
write down.  It contains the same two types of terms,
\eqn{\dfoursg}{\L_D\ =\ \int \dfourth E\,( N^+ N + H_1^+ H_1 + H_2^+
H_2 + ...) \ ,}
and
\eqn{\dtwosg}{\L_F\ =\ \int \dtwoth \E\,( M_N N^2 +
\lambda N^3 + \mu H_1 H_2 + g N H_1 H_2 + ...)\ +\ h.c.\ ,}
where the dots denote nonrenormalizable terms suppressed
by powers of $M_P$.  In these expressions, $E$ and $\E$ are
superspace densities that contain the graviton and gravitino,
as well as the supergravity auxiliary fields $M$ and $b_a$.

The equation of motion for the auxiliary field $M$ contains
a term like
\eqn{\eomM}{M\ \sim\ W/M^2_P + ...\ ,}
where $W$ is the superspotential, which
contains contributions from the visible {\it and}
invisible worlds.  If $\langle W \rangle \sim M_WM^2_P$,
as in hidden-sector models,
then $\langle M \rangle \sim M_W$.  The
vacuum expectation value of $M$ breaks supersymmetry and
communicates the supersymmetry breaking from the invisible
to the visible world.

Because of the nonvanishing value of $\langle M \rangle$,
the superspace densities $E$ and $\E$
are effectively spurions \ref{\GirGri}
{L.~Girardello and M.~Grisaru, Nucl. Phys. B194 (1982) 65.}.
In other words, they induce explicit supersymmetry breaking
into the effective Lagrangian of the visible world.  Indeed,
substituting
\eqn{\spurion}{\eqalign{
\langle E \rangle\ &\sim\ 1 +  \theta\theta \langle M^*\rangle +
\btheta\btheta \langle M \rangle + \theta\theta\btheta\btheta
\langle M^*M \rangle \cr
\ &\sim\ 1 +  \theta\theta M_W + \btheta\btheta M_W +
\theta\theta\btheta\btheta
M^2_W \cr
\langle \E \rangle \ &\sim\ 1 +  \theta\theta \langle M^*\rangle \cr
\ &\sim\ 1 +  \theta\theta M_W\  \cr}}
into \dfoursg\ and \dtwosg, one finds the usual soft breaking
terms (scalar masses and holomorphic bilinear and trilinear
couplings) \GirGri.

\newsec{Nonrenormalizable Couplings}

In a supergravity theory, the effective Lagrangian will
contain a tower of nonrenormalizable operators, suppressed
by appropriate powers of the Planck mass $M_P$.  Because
\321 is unbroken above $M_W$, the higher-dimensional
operators must be \321-invariant.  Such operators
include
\eqn{\nhonehone}{{1\over M_P} \int \dfourth E\,N H^+_1 H_1
\ \sim\ {M^2_W \over M_P} \int d^4x\ n h_1^* h_1 + ...}
\eqn{\nhtwohtwo}{{1\over M_P} \int \dfourth E\,N H^+_2 H_2
\ \sim\ {M^2_W \over M_P} \int d^4x\ n h_2^* h_2 + ...}
and
\eqn{\dhdh}{ {1\over M_P} \int \dfourth E\,D^\alpha H_1 D_\alpha H_2
\ \sim\ {M^2_W \over M_P} \int d^4x\ \chi_1\chi_2 + ...}
where the dots denote additional terms that come from the
lower components of $\langle E \rangle$.  Note
that these operators include arbitrary trilinear couplings and
explicit masses for the fermionic partners of the Higgs fields.
They are hard supersymmetry breakings \GirGri, although their
mass parameters are small, of order $M^2_W/M_P$.

In general, such hard supersymmetry breakings can
destabilize the hierarchy.  To see whether they do, one must first
identify the dangerous diagrams.  The usual formula for the
superficial degree of divergence for a $D$-type superspace diagram
can be readily generalized to include nonrenormalizable operators.
It becomes
\eqn{\ddiv}{D \ \le \ 2 - E_c  - P_c + \sum_d d V_d\ ,}
where $E_c$ denotes the number of chiral external legs, $P_c$
represents the number of chiral $\langle \Phi\Phi \rangle$
propagators, and $V_d$ denotes the number of nonrenormalizable
operators suppressed by $(1/M_P)^d$.

Using this formula, it is easy to see that a given diagram will
be proportional to
\eqn{\diag}{\Lambda^D\ \prod_d\ \bigg({1 \over M_P}\bigg)^{d V_d}\ .}
If we take the cutoff $\Lambda$ to be of order $M_P$, this
reduces to
\eqn{\diagg}{M_P^{2-E_c-P_c}\ .}
Equation \diagg\ tells us that a superspace tadpole diagram can be at
most linearly divergent, while the divergence associated with a
superspace two-point function can be at most logarithmic.

The formula \diagg\ tells us immediately that models without
singlet superfields are automatically safe from destabilizing
divergences.  Gauge (and/or discrete symmetry) invariance
forbids the appearance of tadpoles, so no divergences appear
that can possibly spoil the hierarchy.  In particular, the minimal
supersymmetric standard model is safe \ref{\HallRandall}{L.~Hall
and L.~Randall, Phys. Rev. Lett. 65 (1990) 2939.}.

In contrast, hidden-sector models with (visible) singlet
superfields are potentially unstable.  This is easy to check using
the above example.  Indeed, each of the couplings \nhonehone,
\nhtwohtwo\ and \dhdh\ can give rise to a divergent tadpole,
as illustrated in Figure 1.  Each of the diagrams gives rise
to an operator of the form
\eqn{\tad}{{\Lambda^2 \over M_P}\,\int \dfourth E N\
\ \sim\ {M^2_W \Lambda^2\over M_P}\,\int d^4x\ n\ +\
{M_W \Lambda^2\over M_P}\,\int d^4x\ F_n\ .}
This operator induces a vacuum expectation value of order
$M^2_W \Lambda^2/M_N M_P$ for $F_n$.  For singlet masses
$M_N \lsim \Lambda^2/M_P$, the scale of
supersymmetry breaking in the visible world is greater than
$M_W$.  This will drive up the masses of the
superparticles and destabilize the hierarchy.

The destabilization can be seen explicitly by noting that
$\langle F_n \rangle$ induces a mass for the Higgs multiplet
of order $M_W\Lambda/\sqrt{M_N M_P}$,
\eqn{\mass}{\eqalign{
\int \dtwoth N H_1 H_2\ &\rightarrow \ \int d^4x\ \langle F_n
\rangle h_1 h_2\ + \ ...\cr
  & \sim \ (M^2_W \Lambda^2/ M_N M_P) h_1 h_2\ +\ ...\cr}}
If $\Lambda \sim M_P$ and $M_N \sim M_W$, the Higgs mass becomes
$M_H \sim \sqrt{M_W M_P}$.

Clearly, for the hierarchy to be stable, eqn.~\mass
\ implies that the cutoff $\Lambda$ must be taken to be
less than $\sqrt{M_N M_P}$.
This places an upper bound on the scale of new physics in
hidden-sector supersymmetric theories with visible
singlets.  For $M_N \sim M_W$, the upper bound becomes
$\Lambda \lsim \sqrt{M_WM_P}$, the scale
of supersymmetry breaking in the hidden sector.

\newsec{Conclusion}

In this letter we have shown that gravitational radiative corrections
can destabilize the hierarchy in hidden-sector supergravity-coupled
supersymmetric theories with visible singlets.
This is true whether or not the theory is unified at
some scale $M_X \gg M_W$.  The minimal supersymmetric standard model
is safe from such destabilizing divergences.

We would like to thank M. Einhorn for asking whether
gravitational radiative corrections can destabilize the hierarchy.
We would also like to thank K. Matchev, R. Zhang
and D. Nemeschansky for helpful discussions during the
course of this work, and U.~Ellwanger for drawing
our attention to Ref. \Ellwanger.
Finally, J.B. would like to express his
appreciation to the Aspen Center for Physics for providing a
stimulating atmosphere in which to complete this work.

\vfill\eject

\nfig{\fifone}{One-loop superspace tadpole diagrams for $N$, induced
by the operators \nhonehone\ -- \dhdh.}

\vskip-1in\immediate\closeout\ffile{\parindent40pt
\baselineskip14pt\noindent{\bf Figure Captions}\medskip
\escapechar=` \input figs.tmp\vfill\eject}
\footatend\vfill\eject\immediate\closeout\rfile\writestop
\baselineskip=14pt\noindent{\bf References}\bigskip{\frenchspacing%
\parindent=20pt\escapechar=` \input
refs.tmp\vfill\eject}\nonfrenchspacing
\end